\documentclass[sigconf]{acmart}

\usepackage{booktabs} 
\usepackage{multirow}
\usepackage{soul}
\usepackage{array}
\usepackage{verbatim}
\usepackage{threeparttable}
\usepackage{tabularx}
\newcolumntype{Y}{>{\centering\arraybackslash}X}
\usepackage{enumitem}
\usepackage{makecell}
\usepackage{tabu}
\usepackage{color}
\usepackage{subcaption}

\usepackage[para]{footmisc} 

\usepackage[textsize=scriptsize,backgroundcolor=green,disable]{todonotes}
\title{Mix and Match: Collaborative Expert-Crowd Judging for Building Test Collections Accurately and Affordably}
 
\begin{document}

\author{Mucahid Kutlu}
\affiliation{%
  \institution{Qatar University}
}
\email{mucahidkutlu@qu.edu.qa}

\author{Tyler McDonnell}
\affiliation{%
  \institution{University of Texas at Austin}
}
\email{tyler@cs.utexas.edu}

\author{Aashish Sheshadri}
\affiliation{%
  \institution{PayPal}
}
\email{aashish.sheshadri@gmail.com}

\author{Tamer Elsayed}
\affiliation{%
  \institution{Qatar University}
}
\email{telsayed@qu.edu.qa}

\author{Matthew Lease}
\affiliation{%
  \institution{University of Texas at Austin}
}
\email{ml@utexas.edu}


\begin{abstract}
Crowdsourcing offers an affordable and scalable means to collect relevance judgments for IR test collections. However, crowd assessors may show higher variance in judgment quality than trusted assessors. In this paper, we investigate how to effectively utilize both groups of assessors in partnership. We specifically investigate how agreement in judging is correlated with three factors: relevance category, document rankings, and topical variance. Based on this, we then propose two collaborative judging methods in which a portion of the document-topic pairs are assessed by in-house judges while the rest are assessed by crowd-workers. Experiments conducted on two TREC collections show encouraging results when we distribute work intelligently between our two groups of assessors.
\end{abstract}


\maketitle

\section{Introduction}

Crowdsourcing platforms such as Mechanical Turk (MTurk), provide a new avenue for scalable, low-cost collection of relevance judgments for constructing Information Retrieval (IR) test collections \cite{alonso2009can}. However, quality of crowd judgments appears more variable than the traditional practice use of in-house, trusted personnel. Consequently, researchers have explored various techniques for quality assurance with crowdsourcing, including design to discourage cheating \cite{kittur2008}, predicting correctness of future contributions~\cite{jung2014predicting}, and requesting rationales behind their judgments~\cite{mcdonnell2016relevant}. Prior work has also studied whether crowd-workers are suitable for particular IR tasks, such as domain-specific search~\cite{clough2013examining} and e-discovery~\cite{MEET:MEET14504503126}. 


In this work, we hypothesize that crowd judges may be better suited to judge some documents than others for relevance. If we could effectively distinguish such documents, then we could effectively route only those appropriate documents to the crowd for judging, while restricting our more limited and expensive trusted judges to remaining documents \cite{Nguyen15-hcomp}. Since our goal is to utilize trusted judges only for the documents that we believe the crowd is likely to label incorrectly, we investigate three broad factors which may correlate with agreement in judging: relevance category, document rankings, and topical variance. This builds on a long and storied history of research study on disagreement in relevance judging \cite{saracevic2008effects,lesk1969interactive,funk1983indexing,burgin1992variations,Voorhees2000697,sormunen2002liberal,bailey2008relevance,carterette2010effect,kazai2012analysis,chouldechova2013differences,al2014qualitative}.

Following this, we evaluate two collaborative judging approaches. The first {\em oracle} approach 
uses knowledge of disagreement for each topic to prioritize high disagreement topics for trusted judging.
 The second, practical approach focused on document importance rather than expected disagreement, using expensive trusted judges to judge those highly ranked documents which most greatly impact rank-based evaluation metrics. In particular, we use statAP method~\cite{pavlu2007practical}'s weighting function to prioritize highly-ranked documents for trusted judging. We compare both approaches to a random document ordering baseline.

We report experiments using two TREC test collections for which both crowd and NIST (i.e., trusted) judgments are available. Results show that collaborative judging offers a promising method to leverage the crowd in combination with trusted judges for accurate and affordable building of IR test collections.

\section{Disagreement in Judgments}
\label{sec_disagreement}

To better understand on which topic-document pairs we might expect to see judging disagreement, we investigate three broad factors which may correlate with such disagreement: relevance category, document rankings, and topical variance.

\subsection{Test Collections}

We use two test collections to investigate judging disagreement between crowd-workers and NIST assessors, and to conduct rank correlation experiments using collaborative judging.

\begin{table*}[!htb]
\centering
\caption{Confusion Matrices for Crowd (Cr) vs. NIST Judgments in MQ'09 and WT'14 Test Collections. 'R'  represents relevant judgments and 'NR' represents not-relevant judgments. Bold indicates agreement. }
\label{tab_confusion_matrices}
\begin{tabularx}{0.9\textwidth}{lYY|YY||Y@{\extracolsep{12pt}}YY|YY||Y@{\extracolsep{12pt}}}
\hline\noalign{\smallskip}
 & \multicolumn{5}{c}{\bf \textit{MQ'09}} & \multicolumn{5}{c}{\bf \textit{WT'14} } \\ 
\cline{2-6}\cline{7-11}\noalign{\smallskip} 
   & \multicolumn{2}{c}{\bf \textit{ Majority Voting}} & \multicolumn{2}{c}{\bf \textit{Dawid-Skene}} & & \multicolumn{2}{c}{\bf \textit{ Majority Voting}} & \multicolumn{2}{c}{\bf \textit{Dawid-Skene}} \\
& \multicolumn{2}{c}{\bf 65\%} & \multicolumn{2}{c}{\bf 70\%} & & \multicolumn{2}{c}{\bf 80\%} & \multicolumn{2}{c}{\bf 81\%} \\
   \cline{2-3}\cline{4-5}\cline{7-10}\noalign{\smallskip} 
    & Cr-R&Cr-NR &Cr-R & Cr-NR&  \bf Total & Cr-R&Cr-NR&Cr-R&Cr-NR &   \bf Total \\ \hline
   NIST-R 	& \textbf{44\%} &10\% &\textbf{41\%} &13\% 	&54\% 	&\textbf{39\%} &6\%	& \bf 37\% &8\%   &45\%  \\
    NIST-NR & 25\% &\textbf{21\%} &17\% &\textbf{29\%} 	&46\% 	&14\% &\textbf{41\%}	&11\% & \bf 44\%  &55\%  \\
    \hline
    \bf Total 		& 69\% &31\% &58\% &42\% 	&100\%	&53\% &47\% &48\%  &52\% &100\% \\
\noalign{\smallskip}\hline
  \end{tabularx}
	\end{table*}

\textbf{Million Query Track 2009 (MQ'09)} \cite{carterette2009million}. 100 MQ'09 topics, as well as the ClueWeb09  collection\footnote{\url{http://lemurproject.org/clueweb09/}} were reused in the TREC Relevance Feedback (RF'09) Track ~\cite{Buckley10-notebook}. Because RF'09 participating systems retrieved additional documents not judged for MQ'09, additional relevance judgments were collected for the track via MTurk. These judgments were also used for the subsequent TREC Crowdsourcing Track\footnote{\url{https://sites.google.com/site/treccrowd/}} and made freely available. Beyond judging new documents, 3,277 documents already judged by NIST were also re-judged as part of quality assurance during data collection. As a result, we have 20,535 crowd judgments which we can measure agreement with NIST. We also evaluate 35 system runs submitted to MQ'09 using these crowd judgments vs. NIST judgments, measuring rank correlation.

\textbf{Web Track 2014 (WT'14)} \cite{collins2015trec}. Recently, a new crowd-judgment collection has been released~\cite{tanyahcomp2018,sigir18}. 100 NIST-judged documents for each of 50 WT'14 topics were selected by statAP~\cite{pavlu2007practical}'s sampling method. MTurk judgments for these documents were collected via \cite{mcdonnell2016relevant,mcdonnell2017ijcai}'s rationale method. In total, 25,099 MTurk judgments  for 5000 documents were  collected (i.e., roughly 5 judgments per document).
We evaluate 29 system runs submitted to WT'14 using these crowd judgments vs.\ NIST judgments, measuring rank correlation.

For both test collections, we reduce graded relevance judgments to being binary and report two different methods for aggregating them: majority voting (MV) and Dawid-Skene (DS)~\cite{dawid1979maximum}. Whereas MV performs unweighted voting, DS performs weighted voting based on unsupervised individual reliability estimates. Agreement statistics in {\bf Table \ref{tab_confusion_matrices}} show the WT'14 crowd is 10-15\% more accurate than the MQ'09 crowd, {\em wrt.} NIST judgments as the ``gold standard'' (80\% vs. 65\% in MV and 81\% vs. 70\% in DS). We also see that DS performs much better on MQ'09 (where it evidences more variability in crowd assessor reliability) than on WT'14, whose crowd demonstrates less variability.



\subsection{Agreement vs.\ Relevance Category}

Is there more disagreement on documents judged by NIST to be relevant or non-relevant\todo{by collapsing NIST graded judgments to binary, we lose an opportunity to see more disagreement on borderline middle relevance grades.}? We might expect higher agreement for clear-cut cases of relevance/non-relevance, and higher disagreement for boundary cases \cite{lesk1969interactive,Voorhees2000697}. {\bf Table~\ref{tab_confusion_matrices}} shows confusion matrices for both test collections and aggregation methods. For all settings, we observe that crowd judgments show higher agreement with NIST assessors on NIST-judged relevant documents than on non-relevant ones. Assuming DS aggregation, we see that crowd judgments  agree with NIST on $\frac{41\%}{54\%}=76\%$ (MQ'09) and $\frac{37\%}{45\%}=82\%$ (WT'14) of NIST-judged relevant documents. On non-relevant documents,  crowd judgments  agree with NIST to a lesser degree: $\frac{29\%}{46\%}=63\%$ (MQ'09) and $\frac{44\%}{55\%}=80\%$ (WT'14). Higher agreement 
on NIST-judged relevant documents 
suggests that when in doubt which way to lean, non-expert judges may be more liberal in judging documents as relevant \cite{sormunen2002liberal}.

\begin{figure*}[!htb]
\centering
	\begin{subfigure}{0.45\textwidth}
		\includegraphics[width=0.9\textwidth]{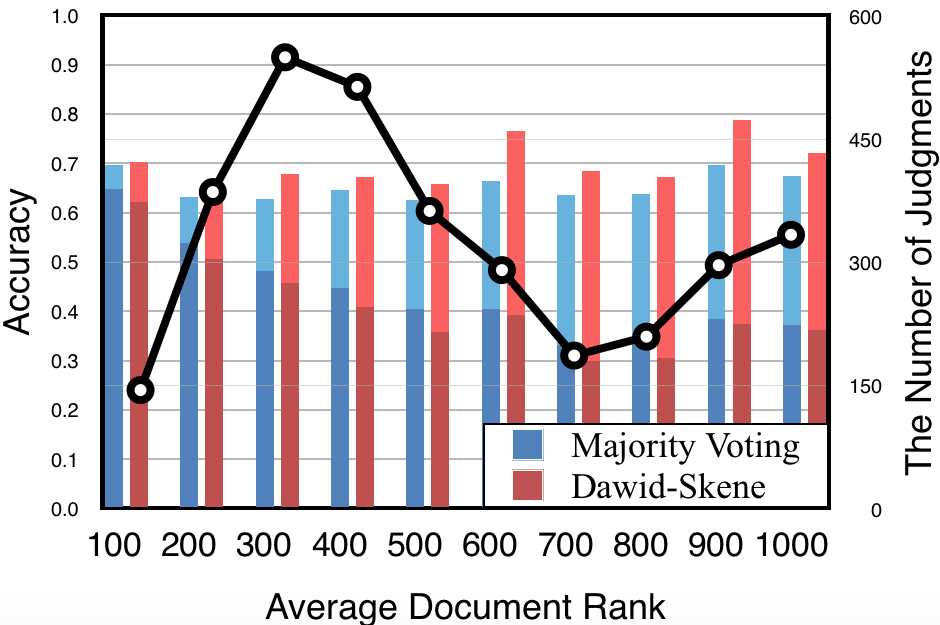} 
  		\caption{MQ'09 }
  \end{subfigure}
   \begin{subfigure}{0.45\textwidth}
		\includegraphics[width=0.9\textwidth]{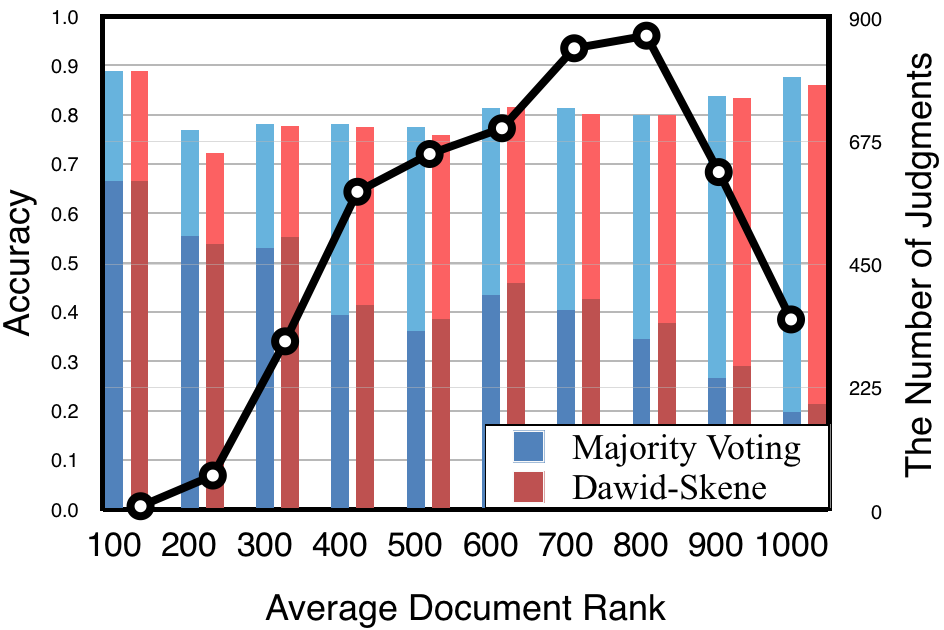} 
  		\caption{WT'14}
  \end{subfigure}
    \caption{Accuracy of crowd judgments vs.\ average document rank. Bars show accuracy and are shaded: lower, darker regions represent the ratio of true positives, while higher, lighter regions represent the ratio of true negatives. The black line shows the number of judgments per bin.}
    \label{fig_average_rank}
\end{figure*}

\subsection{Agreement vs.\ Document Rank}

We next explore how disagreement is correlated with document rankings. Following the same logic discussed above regarding document relevance, we might expect clearly relevant documents to retrieved at very high ranks, clearly non-relevant documents to be retrieved at low ranks, and borderline documents (leading to judging disagreement) retrieved at more middling ranks \cite{lesk1969interactive,Voorhees2000697}. To inspect this here, we compute the average rank of each judged document across all submitted runs for each track. We treat all documents retrieved at rank $\geq$ 1000 as having rank 1000, then bin documents into 10 groups based on their average ranks, using a fixed interval size of 100 ranks. Finally, we compute agreement statistics for each bin for NIST vs.\ crowd-workers. 

{\bf Figure~\ref{fig_average_rank}} shows the accuracy of the 10 groups over the two collections.
In MQ'09, the accuracy for the first group (i.e., the average document rank $\leq 100$) is higher than the accuracy when the average document rank is between 200-500 in both aggregation methods. Regarding the results for WT'14, we observe a more clear pattern: 
The accuracy is noticeably higher in the first group. The accuracy for the second group is the lowest among other groups and then the accuracy increases  gradually  as the average document rank increases. 

\subsection{Agreement Across Topics}
\label{sec_topic}

Because an individual with a given information need knows best what they are and are not looking for \cite{chouldechova2013differences}, NIST typically utilizes the same individual to both develop a topic (description) and perform judging for that topic. While written topic descriptions are useful, they are never complete, and so secondary assessors (be they NIST \cite{Voorhees2000697} or crowd  \cite{alonso2009can}) have less information to go on when judging relevance of someone else's topic. This naturally leads to disagreement. While even NIST assessors are known to often disagree~\cite{Voorhees2000697},  crowd judging introduces further variability. For example, former intelligence analysts working as NIST assessors may share a common geographic, cultural, and knowledge background, suggesting a consistent bias. Crowd workers may be far more diverse. 

\begin{figure*}
\centering
	\begin{subfigure}{0.45\textwidth}
		\includegraphics[width=0.9\textwidth]{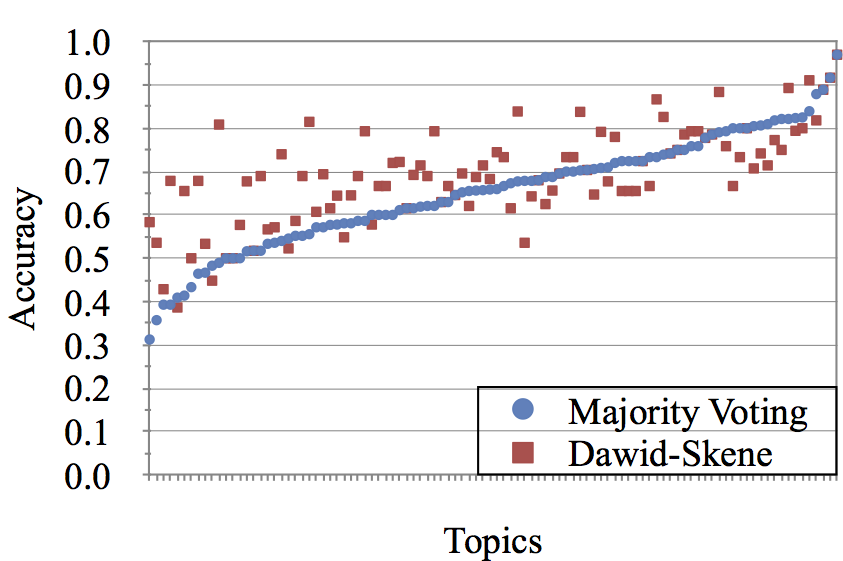} 
  		\caption{MQ'09}
  \end{subfigure}
   \begin{subfigure}{0.45\textwidth}
		\includegraphics[width=0.9\textwidth]{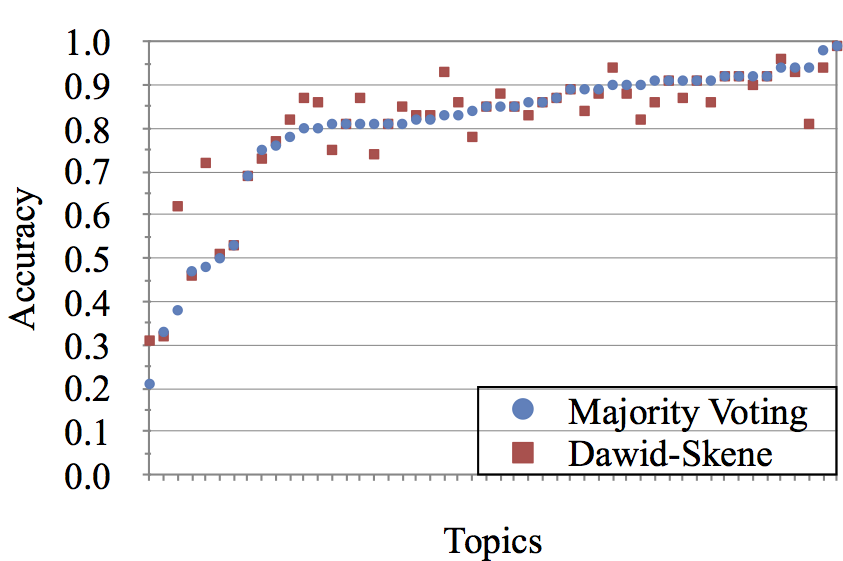} 
  		\caption{WT'14}
  \end{subfigure}
    \caption{Distribution of crowd judging agreement with NIST across topics for MQ'09 and WT'14. Topics are ordered left-to-right by increasing Majority Voting aggregation accuracy.}
    \label{fig_accuracy_per_topic}
\end{figure*}

{\bf Figure~\ref{fig_accuracy_per_topic}} shows the distribution of judging agreement across topics for MQ'09 and WT'14.  
With MV aggregation, the standard deviation across topics is 0.13 (MQ'09) and 0.17 (WT'14). However, with DS aggregation, the standard deviation slightly tightens to 0.11 (MQ'09) and 0.15 (WT'14). Interestingly, the standard deviation is actually higher in WT'14, despite its crowd judgments having higher accuracy. Regardless, we clearly do observe large variability in judging agreement across different topics in both collections, suggesting the importance of modeling topical factors in order to accurately predict assessor disagreement \cite{Webber:2012:AAD:2396761.2396781,Chandar:2013:DFP:2484028.2484161,Sheshadri-thesis14}.



\section{Collaborative Judging}
\label{sec_collaborative_judging}

Thus far we have seen (1) greater agreement on NIST-judged relevant documents, potentially due to layman crowd judges being permissive in judging relevance when in doubt \cite{sormunen2002liberal}; (2) greater agreement on documents ranked very high or low \cite{lesk1969interactive,Voorhees2000697}; and (3) high variance in agreement across topics. Based on this, we evaluate two simple methods for collaborative judging: a practical method prioritizing highly-ranked documents for trusted judging because of their significant impact on ranking metrics, and an oracle method which prioritizes documents based on knowledge of disagreement for each topic.



\subsection{Descending Rank Based Ordering (DRBO)} 
Documents at higher ranks more significantly impact rank-based evaluation metrics of IR system performance (e.g., MAP). Therefore, an intuitive method for collaborative judging would be to assign these more important documents to trusted judges. 

Specifically, we calculate the weight of each document-topic pair using statAP method~\cite{pavlu2007practical}'s weighting function, which assigns higher scores to documents at higher ranks. Subsequently, we rank the documents based on their weights in descending order, for each topic. The first $K$ documents of each topic are assigned to the trusted judges, while the rest are judged by crowd workers. 

\subsection{Oracle Topic-Based Scheduling (Oracle TBS)}
As discussed in Section~\ref{sec_topic}, judging disagreement exhibits high variance across topics. Therefore, quality of judgments might be improved by assigning to trusted judges those topics for which the most disagreement is expected. However, it is challenging to predict which topics are easier for crowd workers to judge \cite{Webber:2012:AAD:2396761.2396781,Chandar:2013:DFP:2484028.2484161,Sheshadri-thesis14}. We leave this for future work and instead pursue a preliminary oracle model as a proof-of-concept which  perfectly predicts judging agreement for each topic. Using this oracle, we then prioritize documents for expert judging starting with the lowest agreement topics first. Documents for the same topic are ordered randomly. 

\begin{figure*}[ht]
\centering
  \begin{subfigure}{7.0cm}
		\includegraphics[width=7.0cm]{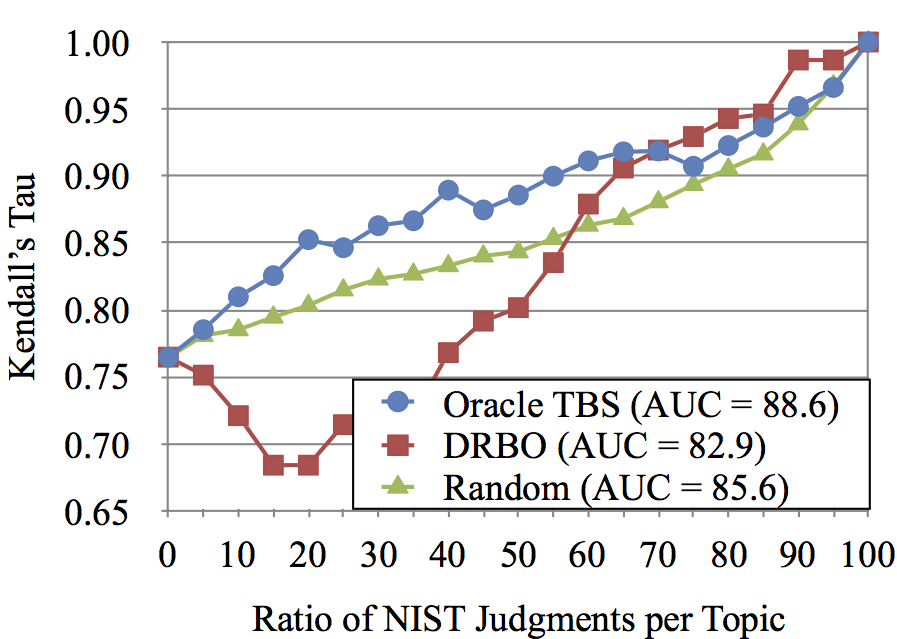} 
  		\caption{MQ'09 - Majority Voting}
  \end{subfigure}
  \begin{subfigure}{7.0cm}
		\includegraphics[width=7.0cm]{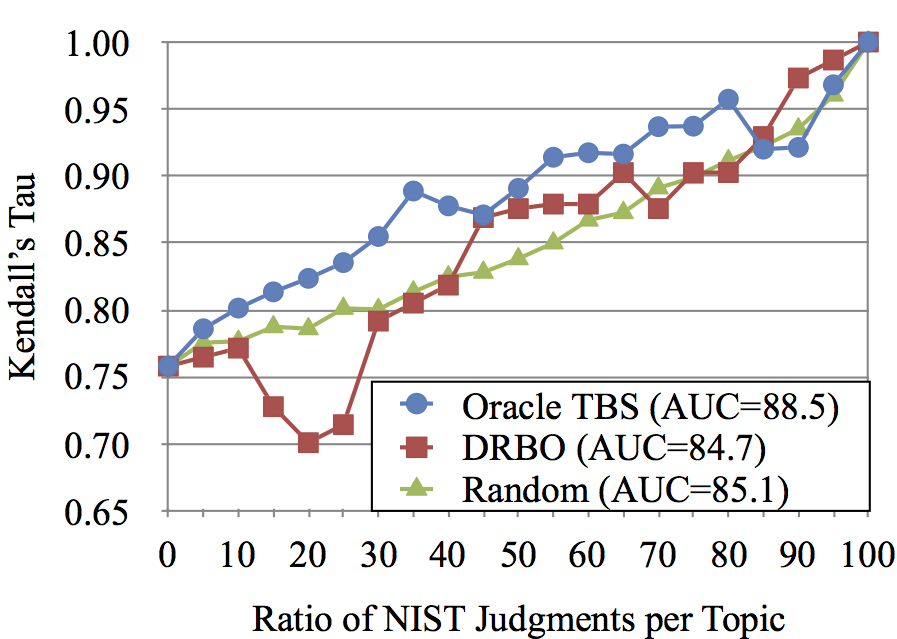} 
  		\caption{MQ'09 - Dawid-Skene}
  \end{subfigure}
  	\begin{subfigure}{7.0cm}
		\includegraphics[width=7.0cm]{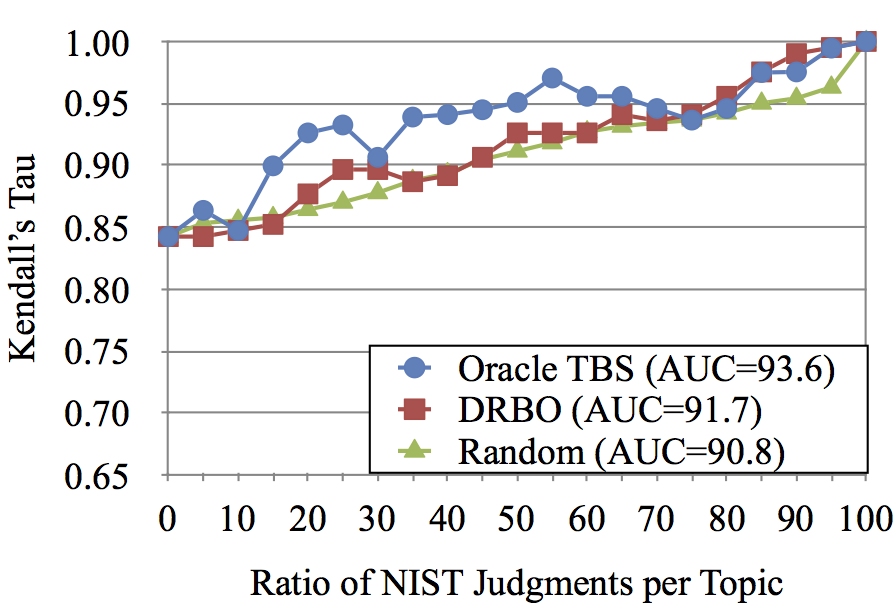} 
  		\caption{WT'14 - Majority Voting}
  \end{subfigure}
   \begin{subfigure}{7.0cm}
		\includegraphics[width=7.0cm]{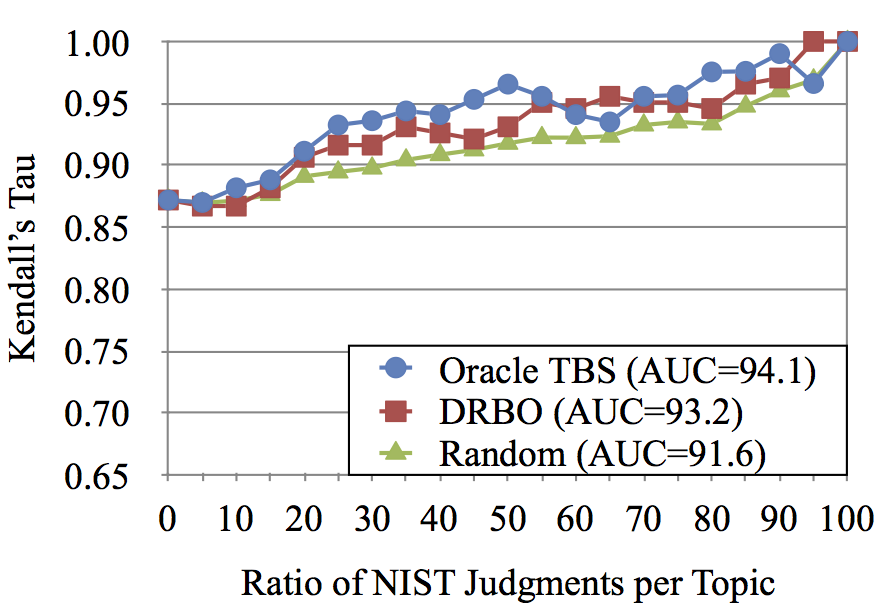} 
  		\caption{WT'14 - Dawid-Skene}
  \end{subfigure}
    \caption{Comparing cost vs.\ correlation of Oracle TBS, DRBO, and Random methods for collaborative NIST and crowd judging.}.
    \label{fig_collaborative_judging}
\vspace{-15pt}
\end{figure*}

\section{Experiments}\label{sec_experiments}

In this section, we report experiments to compare the proposed collaborative judging methods on our test collections. We also report the \emph{random} method,  which  assigns randomly-selected $K$ documents  to the trusted personnel for each topic, as a baseline.

Because the number of judged documents per topic greatly varies, we vary the ratio of NIST judgments used per topic, instead of using a fixed constant.  Given our incomplete judgments, we evaluate IR systems using \emph{bpref} \cite{buckley2004retrieval}, which ignores the documents for which no judgment is available. We adopt Soboroff's corrected bpref formulation \cite{soboroff2006dynamic}, as implemented in {\tt trec\_eval}\footnote{\url{http://trec.nist.gov/trec_eval/}}. We assume that ground-truth ranking of systems comes from ranking systems based on their bpref scores using NIST judgments. We  calculate Kendall's $\tau$ 
to measure correlation between the ground-truth ranking and the ranking induced by collaborative judging. We run the random method 50 times and the Oracle TBS method 10 times and report average Kendall's $\tau$. By convention, $\tau=0.9$ is assumed to constitute an acceptable correlation level for reliable IR evaluation \cite{Voorhees2000697}.

Results are shown in \textbf{Figure~\ref{fig_collaborative_judging}}. We report the area-under-curve (AUC) in the Figure to have a simple summarization of the results. The results suggest several observations. Firstly, the Oracle TBS method achieves the best overall results, reaching $\tau=0.9$ score by assigning 55\% and 15-20\% of the judgments to the trusted personnel in MQ'09 and WT'14, respectively. This suggests that if we could predict which topics will more likely exhibit judging disagreement, then we might maintain NIST quality judging at lower cost through collaborative judging. Secondly, DRBO consistently outperforms the random baseline in WT'14, but not in MQ'09. This may be due to lower quality crowd judgments in MQ'09 (See Table~\ref{tab_confusion_matrices}). With higher quality crowd judgments, however, DRBO seems to be a  simple and effective method. Overall, our results suggest that collaborative judging is a promising method to efficiently build high-quality test collections.

\section{Conclusion and Future Work}

In this work, we investigated when crowd workers disagree with NIST assessors and proposed  one oracle and one practical collaborative judging approach. Based on our experiments conducted on two different test collections, we offer several observations. 

First, higher agreement with NIST on documents NIST judges to be relevant suggests that when in doubt, layman crowd judges may be more liberal in erring on the side of judging unsure documents as relevant~\cite{sormunen2002liberal}. Secondly, we reaffirm prior work's finding of greater judging agreement at very high and low ranks, suggesting documents whose relevance is not borderline \cite{lesk1969interactive,Voorhees2000697}. Thirdly, we do see high variance in agreement across topics, suggesting further confirmation of judging differences between primary and secondary assessors~\cite{Voorhees2000697,chouldechova2013differences,al2014qualitative}. 

In regard to collaborative judging, the oracle predicting assessor disagreement also achieved the highest rank correlation, suggesting that a model which could effectively predict judging disagreement \cite{Chandar:2013:DFP:2484028.2484161,Webber:2012:AAD:2396761.2396781,Sheshadri-thesis14} could be usefully applied toward collaborative judging. Alternatively, the DRBO approach proposed here based on statAP \cite{pavlu2007practical} often outperformed the random baseline across collections and aggregation algorithms, and especially with higher quality crowd judgments, suggesting its promise particularly when crowd judgments are collected carefully.

In the future, we plan to use and extend disagreement prediction modeling \cite{Chandar:2013:DFP:2484028.2484161,Webber:2012:AAD:2396761.2396781,Sheshadri-thesis14} to further improve collaborative judging. Since our current DRBO model prioritizes highly-ranked documents without considering expected disagreement, it would be of further interest to integrate both aspects into a single, coherent strategy for collaborative judging.

\bibliographystyle{ACM-Reference-Format}
\bibliography{references} 


\begin{thebibliography}{00}


\ifx \showCODEN    \undefined \def \showCODEN     #1{\unskip}     \fi
\ifx \showDOI      \undefined \def \showDOI       #1{#1}\fi
\ifx \showISBNx    \undefined \def \showISBNx     #1{\unskip}     \fi
\ifx \showISBNxiii \undefined \def \showISBNxiii  #1{\unskip}     \fi
\ifx \showISSN     \undefined \def \showISSN      #1{\unskip}     \fi
\ifx \showLCCN     \undefined \def \showLCCN      #1{\unskip}     \fi
\ifx \shownote     \undefined \def \shownote      #1{#1}          \fi
\ifx \showarticletitle \undefined \def \showarticletitle #1{#1}   \fi
\ifx \showURL      \undefined \def \showURL       {\relax}        \fi
\providecommand\bibfield[2]{#2}
\providecommand\bibinfo[2]{#2}
\providecommand\natexlab[1]{#1}
\providecommand\showeprint[2][]{arXiv:#2}

\bibitem[\protect\citeauthoryear{Al-Harbi and Smucker}{Al-Harbi and
  Smucker}{2014}]%
        {al2014qualitative}
\bibfield{author}{\bibinfo{person}{Aiman~L Al-Harbi} {and}
  \bibinfo{person}{Mark~D Smucker}.} \bibinfo{year}{2014}\natexlab{}.
\newblock \showarticletitle{A qualitative exploration of secondary assessor
  relevance judging behavior}. In \bibinfo{booktitle}{{\em Proceedings of the
  5th Information Interaction in Context Symposium}}. ACM,
  \bibinfo{pages}{195--204}.
\newblock


\bibitem[\protect\citeauthoryear{Alonso and Mizzaro}{Alonso and
  Mizzaro}{2009}]%
        {alonso2009can}
\bibfield{author}{\bibinfo{person}{Omar Alonso} {and} \bibinfo{person}{Stefano
  Mizzaro}.} \bibinfo{year}{2009}\natexlab{}.
\newblock \showarticletitle{Can we get rid of TREC assessors? Using Mechanical
  Turk for relevance assessment}. In \bibinfo{booktitle}{{\em Proceedings of
  the SIGIR 2009 Workshop on the Future of IR Evaluation}},
  Vol.~\bibinfo{volume}{15}. \bibinfo{pages}{16}.
\newblock


\bibitem[\protect\citeauthoryear{Bailey, Craswell, Soboroff, Thomas, de~Vries,
  and Yilmaz}{Bailey et~al\mbox{.}}{2008}]%
        {bailey2008relevance}
\bibfield{author}{\bibinfo{person}{Peter Bailey}, \bibinfo{person}{Nick
  Craswell}, \bibinfo{person}{Ian Soboroff}, \bibinfo{person}{Paul Thomas},
  \bibinfo{person}{Arjen~P de Vries}, {and} \bibinfo{person}{Emine Yilmaz}.}
  \bibinfo{year}{2008}\natexlab{}.
\newblock \showarticletitle{Relevance assessment: are judges exchangeable and
  does it matter}. In \bibinfo{booktitle}{{\em SIGIR}}. ACM,
  \bibinfo{pages}{667--674}.
\newblock


\bibitem[\protect\citeauthoryear{Buckley, Lease, and Smucker}{Buckley
  et~al\mbox{.}}{2010}]%
        {Buckley10-notebook}
\bibfield{author}{\bibinfo{person}{Chris Buckley}, \bibinfo{person}{Matthew
  Lease}, {and} \bibinfo{person}{Mark~D. Smucker}.}
  \bibinfo{year}{2010}\natexlab{}.
\newblock \showarticletitle{{Overview of the TREC 2010 Relevance Feedback Track
  (Notebook)}}. In \bibinfo{booktitle}{{\em {TREC Conference Notebook}}}.
\newblock


\bibitem[\protect\citeauthoryear{Buckley and Voorhees}{Buckley and
  Voorhees}{2004}]%
        {buckley2004retrieval}
\bibfield{author}{\bibinfo{person}{Chris Buckley} {and}
  \bibinfo{person}{Ellen~M Voorhees}.} \bibinfo{year}{2004}\natexlab{}.
\newblock \showarticletitle{Retrieval evaluation with incomplete information}.
  In \bibinfo{booktitle}{{\em Proceedings of the 27th annual international ACM
  SIGIR conference on Research and development in information retrieval}}. ACM,
  \bibinfo{pages}{25--32}.
\newblock


\bibitem[\protect\citeauthoryear{Burgin}{Burgin}{1992}]%
        {burgin1992variations}
\bibfield{author}{\bibinfo{person}{Robert Burgin}.}
  \bibinfo{year}{1992}\natexlab{}.
\newblock \showarticletitle{Variations in relevance judgments and the
  evaluation of retrieval performance}.
\newblock \bibinfo{journal}{{\em Info.\ Processing \& Management\/}}
  \bibinfo{volume}{28}, \bibinfo{number}{5} (\bibinfo{year}{1992}),
  \bibinfo{pages}{619--627}.
\newblock


\bibitem[\protect\citeauthoryear{Carterette, Pavlu, Fang, and
  Kanoulas}{Carterette et~al\mbox{.}}{2009}]%
        {carterette2009million}
\bibfield{author}{\bibinfo{person}{Ben Carterette}, \bibinfo{person}{Virgiliu
  Pavlu}, \bibinfo{person}{Hui Fang}, {and} \bibinfo{person}{Evangelos
  Kanoulas}.} \bibinfo{year}{2009}\natexlab{}.
\newblock \showarticletitle{Million Query Track 2009 Overview.}. In
  \bibinfo{booktitle}{{\em TREC}}.
\newblock


\bibitem[\protect\citeauthoryear{Carterette and Soboroff}{Carterette and
  Soboroff}{2010}]%
        {carterette2010effect}
\bibfield{author}{\bibinfo{person}{Ben Carterette} {and} \bibinfo{person}{Ian
  Soboroff}.} \bibinfo{year}{2010}\natexlab{}.
\newblock \showarticletitle{The effect of assessor error on IR system
  evaluation}. In \bibinfo{booktitle}{{\em Proceedings of the 33rd
  international ACM SIGIR conference on Research and development in information
  retrieval}}. ACM, \bibinfo{pages}{539--546}.
\newblock


\bibitem[\protect\citeauthoryear{Chandar, Webber, and Carterette}{Chandar
  et~al\mbox{.}}{2013}]%
        {Chandar:2013:DFP:2484028.2484161}
\bibfield{author}{\bibinfo{person}{Praveen Chandar}, \bibinfo{person}{William
  Webber}, {and} \bibinfo{person}{Ben Carterette}.}
  \bibinfo{year}{2013}\natexlab{}.
\newblock \showarticletitle{Document Features Predicting Assessor
  Disagreement}. In \bibinfo{booktitle}{{\em 36th ACM SIGIR Conference on
  Research and Development in Information Retrieval}}.
  \bibinfo{pages}{745--748}.
\newblock


\bibitem[\protect\citeauthoryear{Chouldechova and Mease}{Chouldechova and
  Mease}{2013}]%
        {chouldechova2013differences}
\bibfield{author}{\bibinfo{person}{Alexandra Chouldechova} {and}
  \bibinfo{person}{David Mease}.} \bibinfo{year}{2013}\natexlab{}.
\newblock \showarticletitle{Differences in search engine evaluations between
  query owners and non-owners}. In \bibinfo{booktitle}{{\em Proceedings of the
  sixth ACM international conference on Web search and data mining}}. ACM,
  \bibinfo{pages}{103--112}.
\newblock


\bibitem[\protect\citeauthoryear{Clough, Sanderson, Tang, Gollins, and
  Warner}{Clough et~al\mbox{.}}{2013}]%
        {clough2013examining}
\bibfield{author}{\bibinfo{person}{Paul Clough}, \bibinfo{person}{Mark
  Sanderson}, \bibinfo{person}{Jiayu Tang}, \bibinfo{person}{Tim Gollins},
  {and} \bibinfo{person}{Amy Warner}.} \bibinfo{year}{2013}\natexlab{}.
\newblock \showarticletitle{Examining the limits of crowdsourcing for relevance
  assessment}.
\newblock \bibinfo{journal}{{\em IEEE Internet Computing\/}}
  \bibinfo{volume}{17}, \bibinfo{number}{4} (\bibinfo{year}{2013}),
  \bibinfo{pages}{32--38}.
\newblock


\bibitem[\protect\citeauthoryear{Collins-Thompson, Macdonald, Bennett, Diaz,
  and Voorhees}{Collins-Thompson et~al\mbox{.}}{2015}]%
        {collins2015trec}
\bibfield{author}{\bibinfo{person}{Kevyn Collins-Thompson},
  \bibinfo{person}{Craig Macdonald}, \bibinfo{person}{Paul Bennett},
  \bibinfo{person}{Fernando Diaz}, {and} \bibinfo{person}{Ellen~M Voorhees}.}
  \bibinfo{year}{2015}\natexlab{}.
\newblock \showarticletitle{TREC 2014 web track overview}. In
  \bibinfo{booktitle}{{\em TREC}}.
\newblock


\bibitem[\protect\citeauthoryear{Dawid and Skene}{Dawid and Skene}{1979}]%
        {dawid1979maximum}
\bibfield{author}{\bibinfo{person}{Alexander~Philip Dawid} {and}
  \bibinfo{person}{Allan Skene}.} \bibinfo{year}{1979}\natexlab{}.
\newblock \showarticletitle{Maximum likelihood estimation of observer
  error-rates using the EM algorithm}.
\newblock \bibinfo{journal}{{\em Applied Stat.\/}} (\bibinfo{year}{1979}),
  \bibinfo{pages}{20--28}.
\newblock


\bibitem[\protect\citeauthoryear{Efthimiadis and Hotchkiss}{Efthimiadis and
  Hotchkiss}{2008}]%
        {MEET:MEET14504503126}
\bibfield{author}{\bibinfo{person}{Efthimis~N. Efthimiadis} {and}
  \bibinfo{person}{Mary~A. Hotchkiss}.} \bibinfo{year}{2008}\natexlab{}.
\newblock \showarticletitle{Legal discovery: Does domain expertise matter?}
\newblock \bibinfo{journal}{{\em Proceedings of the American Society for
  Information Science and Technology\/}} \bibinfo{volume}{45},
  \bibinfo{number}{1} (\bibinfo{year}{2008}), \bibinfo{pages}{1--2}.
\newblock
\showISSN{1550-8390}


\bibitem[\protect\citeauthoryear{Funk and Reid}{Funk and Reid}{1983}]%
        {funk1983indexing}
\bibfield{author}{\bibinfo{person}{Mark~E Funk} {and}
  \bibinfo{person}{Carolyn~A Reid}.} \bibinfo{year}{1983}\natexlab{}.
\newblock \showarticletitle{Indexing consistency in MEDLINE.}
\newblock \bibinfo{journal}{{\em Bulletin of the Medical Library
  Association\/}} \bibinfo{volume}{71}, \bibinfo{number}{2}
  (\bibinfo{year}{1983}), \bibinfo{pages}{176}.
\newblock


\bibitem[\protect\citeauthoryear{Goyal, McDonnell, Kutlu, Lease, and
  Elsayad}{Goyal et~al\mbox{.}}{2018}]%
        {tanyahcomp2018}
\bibfield{author}{\bibinfo{person}{Tanya Goyal}, \bibinfo{person}{Tyler
  McDonnell}, \bibinfo{person}{Mucahid Kutlu}, \bibinfo{person}{Matthew Lease},
  {and} \bibinfo{person}{Tamer Elsayad}.} \bibinfo{year}{2018}\natexlab{}.
\newblock \showarticletitle{Your Behavior Signals Your Reliability: Modeling
  Crowd Behavioral Traces to Ensure Quality Relevance Annotations}. In
  \bibinfo{booktitle}{{\em Proceedings of the 6th AAAI Conference on Human
  Computation and Crowdsourcing (HCOMP)}}.
\newblock


\bibitem[\protect\citeauthoryear{Jung, Park, and Lease}{Jung
  et~al\mbox{.}}{2014}]%
        {jung2014predicting}
\bibfield{author}{\bibinfo{person}{Hyun~Joon Jung}, \bibinfo{person}{Yubin
  Park}, {and} \bibinfo{person}{Matthew Lease}.}
  \bibinfo{year}{2014}\natexlab{}.
\newblock \showarticletitle{Predicting Next Label Quality: A Time-Series Model
  of Crowdwork}. In \bibinfo{booktitle}{{\em The 2nd AAAI Conference on Human
  Computation \& Crowdsourcing (HCOMP)}}. AAAI.
\newblock


\bibitem[\protect\citeauthoryear{Kazai, Craswell, Yilmaz, and Tahaghoghi}{Kazai
  et~al\mbox{.}}{2012}]%
        {kazai2012analysis}
\bibfield{author}{\bibinfo{person}{Gabriella Kazai}, \bibinfo{person}{Nick
  Craswell}, \bibinfo{person}{Emine Yilmaz}, {and} \bibinfo{person}{Seyed~MM
  Tahaghoghi}.} \bibinfo{year}{2012}\natexlab{}.
\newblock \showarticletitle{An analysis of systematic judging errors in
  information retrieval}. In \bibinfo{booktitle}{{\em 21st ACM intl.\
  conference on Information and knowledge management (CIKM)}}.
  \bibinfo{pages}{105--114}.
\newblock


\bibitem[\protect\citeauthoryear{Kittur, Chi, and Suh}{Kittur
  et~al\mbox{.}}{2008}]%
        {kittur2008}
\bibfield{author}{\bibinfo{person}{Aniket Kittur}, \bibinfo{person}{Ed~H. Chi},
  {and} \bibinfo{person}{Bongwon Suh}.} \bibinfo{year}{2008}\natexlab{}.
\newblock \showarticletitle{Crowdsourcing User Studies with Mechanical Turk}.
  In \bibinfo{booktitle}{{\em Proceedings of the SIGCHI Conference on Human
  Factors in Computing Systems}} {\em (\bibinfo{series}{CHI '08})}.
  \bibinfo{publisher}{ACM}, \bibinfo{address}{New York, NY, USA},
  \bibinfo{pages}{453--456}.
\newblock
\showISBNx{978-1-60558-011-1}


\bibitem[\protect\citeauthoryear{Kutlu, McDonnell, Barkallah, Elsayed, and
  Lease}{Kutlu et~al\mbox{.}}{2018}]%
        {sigir18}
\bibfield{author}{\bibinfo{person}{Mucahid Kutlu}, \bibinfo{person}{Tyler
  McDonnell}, \bibinfo{person}{Yassmine Barkallah}, \bibinfo{person}{Tamer
  Elsayed}, {and} \bibinfo{person}{Matthew Lease}.}
  \bibinfo{year}{2018}\natexlab{}.
\newblock \showarticletitle{Crowd vs.\ Expert: What Can Relevance Judgment
  Rationales Teach Us About Assessor Disagreement?}
\newblock \bibinfo{journal}{{\em The 41st International ACM SIGIR Conference on
  Research and Development in Information Retrieval\/}} (\bibinfo{year}{2018}).
\newblock


\bibitem[\protect\citeauthoryear{Lesk and Salton}{Lesk and Salton}{1969}]%
        {lesk1969interactive}
\bibfield{author}{\bibinfo{person}{Michael~E Lesk} {and}
  \bibinfo{person}{Gerard Salton}.} \bibinfo{year}{1969}\natexlab{}.
\newblock \showarticletitle{Interactive search and retrieval methods using
  automatic information displays}. In \bibinfo{booktitle}{{\em Proceedings of
  the May 14-16, 1969, spring joint computer conference}}. ACM,
  \bibinfo{pages}{435--446}.
\newblock


\bibitem[\protect\citeauthoryear{McDonnell, Kutlu, Elsayed, and
  Lease}{McDonnell et~al\mbox{.}}{2017}]%
        {mcdonnell2017ijcai}
\bibfield{author}{\bibinfo{person}{Tyler McDonnell}, \bibinfo{person}{Mucahid
  Kutlu}, \bibinfo{person}{Tamer Elsayed}, {and} \bibinfo{person}{Matthew
  Lease}.} \bibinfo{year}{2017}\natexlab{}.
\newblock \showarticletitle{{The Many Benefits of Annotator Rationales for
  Relevance Judgments}}. In \bibinfo{booktitle}{{\em Proceedings of the
  Twenty-Sixth International Joint Conference on Artificial Intelligence
  ({IJCAI-17})}}. AAAI, \bibinfo{pages}{4909--4913}.
\newblock


\bibitem[\protect\citeauthoryear{McDonnell, Lease, Kutlu, and
  Elsayed}{McDonnell et~al\mbox{.}}{2016}]%
        {mcdonnell2016relevant}
\bibfield{author}{\bibinfo{person}{Tyler McDonnell}, \bibinfo{person}{Matthew
  Lease}, \bibinfo{person}{Mucahid Kutlu}, {and} \bibinfo{person}{Tamer
  Elsayed}.} \bibinfo{year}{2016}\natexlab{}.
\newblock \showarticletitle{Why Is That Relevant? Collecting Annotator
  Rationales for Relevance Judgments}. In \bibinfo{booktitle}{{\em 4th AAAI
  Conference on Human Computation and Crowdsourcing (HCOMP)}}.
\newblock


\bibitem[\protect\citeauthoryear{Nguyen, Wallace, and Lease}{Nguyen
  et~al\mbox{.}}{2015}]%
        {Nguyen15-hcomp}
\bibfield{author}{\bibinfo{person}{An~Thanh Nguyen}, \bibinfo{person}{Byron~C.\
  Wallace}, {and} \bibinfo{person}{Matthew Lease}.}
  \bibinfo{year}{2015}\natexlab{}.
\newblock \showarticletitle{{Combining Crowd and Expert Labels using Decision
  Theoretic Active Learning}}. In \bibinfo{booktitle}{{\em {Proceedings of the
  3rd AAAI Conference on Human Computation (HCOMP)}}}.
  \bibinfo{pages}{120--129}.
\newblock


\bibitem[\protect\citeauthoryear{Pavlu and Aslam}{Pavlu and Aslam}{2007}]%
        {pavlu2007practical}
\bibfield{author}{\bibinfo{person}{V Pavlu} {and} \bibinfo{person}{J Aslam}.}
  \bibinfo{year}{2007}\natexlab{}.
\newblock \bibinfo{booktitle}{{\em A practical sampling strategy for efficient
  retrieval evaluation}}.
\newblock \bibinfo{type}{{T}echnical {R}eport}. \bibinfo{institution}{Technical
  report, Northeastern University}.
\newblock


\bibitem[\protect\citeauthoryear{Saracevic}{Saracevic}{2008}]%
        {saracevic2008effects}
\bibfield{author}{\bibinfo{person}{Tefko Saracevic}.}
  \bibinfo{year}{2008}\natexlab{}.
\newblock \showarticletitle{Effects of inconsistent relevance judgments on
  information retrieval test results: A historical perspective}.
\newblock \bibinfo{journal}{{\em Library Trends\/}} \bibinfo{volume}{56},
  \bibinfo{number}{4} (\bibinfo{year}{2008}), \bibinfo{pages}{763--783}.
\newblock


\bibitem[\protect\citeauthoryear{Sheshadri}{Sheshadri}{2014}]%
        {Sheshadri-thesis14}
\bibfield{author}{\bibinfo{person}{Aashish Sheshadri}.}
  \bibinfo{year}{2014}\natexlab{}.
\newblock {\em \bibinfo{title}{{A Collaborative Approach to IR Evaluation}}}.
\newblock \bibinfo{thesistype}{Master's\ thesis}. \bibinfo{school}{Department
  of Computer Science, University of Texas at Austin}.
\newblock
\newblock
\shownote{\url{https://repositories.lib.utexas.edu/handle/2152/25910}.}


\bibitem[\protect\citeauthoryear{Soboroff}{Soboroff}{2006}]%
        {soboroff2006dynamic}
\bibfield{author}{\bibinfo{person}{Ian Soboroff}.}
  \bibinfo{year}{2006}\natexlab{}.
\newblock \showarticletitle{Dynamic test collections: measuring search
  effectiveness on the live web}. In \bibinfo{booktitle}{{\em Proc.\ SIGIR}}.
  \bibinfo{pages}{276--283}.
\newblock


\bibitem[\protect\citeauthoryear{Sormunen}{Sormunen}{2002}]%
        {sormunen2002liberal}
\bibfield{author}{\bibinfo{person}{Eero Sormunen}.}
  \bibinfo{year}{2002}\natexlab{}.
\newblock \showarticletitle{Liberal relevance criteria of TREC-: Counting on
  negligible documents?}. In \bibinfo{booktitle}{{\em Proceedings of the 25th
  annual international ACM SIGIR conference on Research and development in
  information retrieval}}. ACM, \bibinfo{pages}{324--330}.
\newblock


\bibitem[\protect\citeauthoryear{Voorhees}{Voorhees}{2000}]%
        {Voorhees2000697}
\bibfield{author}{\bibinfo{person}{Ellen~M. Voorhees}.}
  \bibinfo{year}{2000}\natexlab{}.
\newblock \showarticletitle{Variations in relevance judgments and the
  measurement of retrieval effectiveness}.
\newblock \bibinfo{journal}{{\em Info.\ Processing \& Management\/}}
  \bibinfo{volume}{36}, \bibinfo{number}{5} (\bibinfo{year}{2000}),
  \bibinfo{pages}{697 -- 716}.
\newblock
\showISSN{0306-4573}


\bibitem[\protect\citeauthoryear{Webber, Chandar, and Carterette}{Webber
  et~al\mbox{.}}{2012}]%
        {Webber:2012:AAD:2396761.2396781}
\bibfield{author}{\bibinfo{person}{William Webber}, \bibinfo{person}{Praveen
  Chandar}, {and} \bibinfo{person}{Ben Carterette}.}
  \bibinfo{year}{2012}\natexlab{}.
\newblock \showarticletitle{Alternative Assessor Disagreement and Retrieval
  Depth}. In \bibinfo{booktitle}{{\em Proc.\ 21st ACM CIKM}}.
  \bibinfo{pages}{125--134}.
\newblock


\end{thebibliography}

\end{document}